\documentclass[journal]{IEEEtran}
\usepackage[english]{babel}
\usepackage[algo2e,ruled,linesnumbered]{algorithm2e}

\usepackage{balance}

\usepackage{import}
\usepackage[]{units}
\usepackage{url}
\usepackage{subfigure}
\usepackage{textcomp}
\usepackage{cite}
\SetKwComment{Comment}{$\triangleright$\ }{}
\usepackage{caption}

\ifCLASSINFOpdf
  \usepackage[pdftex]{graphicx}
\else
\fi

\usepackage[cmex10]{amsmath}

\usepackage[inline]{enumitem}

\usepackage[np]{numprint}
\usepackage{hhline}
\npstyleenglish

\usepackage{amssymb}
\usepackage{xcolor}

\graphicspath{{figures/}}

\def\BibTeX{{\rm B\kern-.05em{\sc i\kern-.025em b}\kern-.08em
    T\kern-.1667em\lower.7ex\hbox{E}\kern-.125emX}}

\usepackage[absolute,showboxes]{textpos}

\TPMargin{5pt}

\begin{document}
	
\renewenvironment{IEEEbiography}[1]
{\IEEEbiographynophoto{#1}}
{\endIEEEbiographynophoto}

\bstctlcite{IEEEexample:BSTcontrol}


%
\title{6G opens up a New Era for Aeronautical Communication and Services}
\author{

\IEEEauthorblockN{Arled Papa\IEEEauthorrefmark{1}, J\"{o}rg von Mankowski\IEEEauthorrefmark{1}, Hansini Vijayaraghavan\IEEEauthorrefmark{1}, Babak Mafakheri\IEEEauthorrefmark{2}, Leonardo Goratti\IEEEauthorrefmark{2}, Wolfgang Kellerer\IEEEauthorrefmark{1}\\}
\IEEEauthorblockA{\IEEEauthorrefmark{1}Chair of Communication Networks, Technical University of Munich\\
\IEEEauthorrefmark{2}Safran Passenger Innovations, We{\ss}ling Germany\\
\IEEEauthorrefmark{1}arled.papa@tum.de,
\IEEEauthorrefmark{1}joerg.von.mankowski@tum.de,
\IEEEauthorrefmark{1}hansini.vijayaraghavan@tum.de,
\IEEEauthorrefmark{2}babak.mafakheri@zii.aero,
\IEEEauthorrefmark{2}leonardo.goratti@zii.aero,
\IEEEauthorrefmark{1}wolfgang.kellerer@tum.de}
}

\maketitle

\begin{abstract}

While 5G delivers high quality services mostly in a two dimensional terrestrial area covering our planet's surface, with 6G we aim at a full exploitation of three dimensions.
In this way, 6G includes all kinds of non-terrestrial networks.
In particular, Unmanned Aerial Vehicles (UAVs), High-Altitude Platforms (HAPs), (self-)flying taxis and civil aircrafts are new additions to already existing satellite networks complementing the cellular terrestrial network.
Their integration to 6G is promising with respect to service coverage, but also challenging due to the so far rather closed systems.
Emerging technology concepts such as Mobile Edge Computing (MEC) and Software-Defined Networking (SDN) can provide a basis for a full integration of aeronautical systems into the terrestrial counterpart.
However, these technologies render the management and orchestration of aeronautical systems complex.
As a step towards the integration of aeronautical communication and services into 6G, we propose a framework for the collection, monitoring and distribution of resources in the sky among heterogeneous flying objects. 
This enables high-performance services for a new era of 6G aeronautical applications.
Based on our aeronautical framework, we introduce emerging application use-cases including Aeronautical Edge Computing (AEC), aircraft-as-a-sensor, and in-cabin networks.

\end{abstract}

\section{Introduction}

While 5G cellular communication networks are being deployed, research started to address the next generation targeting the year 2030.
Whereas 5G has put the focus on machine-type communication, 6G is expected to bring the consumer in the center.
In particular, the goal is to extend the human intelligence in any situation and environment.
In this way, 6G is envisioned to advance the initial efforts of 5G for embracing also networks that have not been in the core focus of terrestrial cellular communications so far. Thus, including Unmanned Aerial Vehicles (UAVs), High-Altitude Platforms (HAPs), (self-)flying taxis and civil aircrafts~\cite{azari2021evolution}. Furthermore, 6G is anticipated to extend ubiquitous networking and service access in a full three-dimensional sphere, where aeronautical and non-terrestrial networks play an important role~\cite{strinati20206g}. Moreover, as presented in~\cite{henry20205g}, 5G networks can cater for mobility speeds of up to $500$\,km/h, which is not suitable for aeronautical communication, yet it could be accommodated with 6G.

The main objective for aeronautical systems is a full integration into the 6G infrastructure to extend the range of applications and use cases.
That entails full incorporation of 6G features into the aeronautical architecture, aiming at achieving end-to-end Quality of Service (QoS).
Emerging 6G applications~\cite{viswanathan2020communications} range from holographic telepresence over tele-operation in medicine, industrial robot swarms and digital twins to self-driving vehicles.
The natural question then arises.
\emph{What are the most important 6G features supporting such applications and can they be mapped to aeronautical networks}?
Most of the emerging 6G technologies include cognitive spectrum sharing, new communication paradigms (e.g., quantum communication, semantic communication), Artificial Intelligence (AI)-driven communication, Mobile Edge Computing (MEC), localization and sensing, extreme performance guarantees, security, privacy and sustainability. In the latter, we show that the aforementioned technologies can also be mapped to aeronautical networks.  

In particular, to leverage the full potential of AI, MEC, localization, sensing, security, privacy and sustainability in the sky an end-to-end integration and cooperation of the different heterogeneous aeronautical systems is necessary to deliver high performance services.
To date, the heterogeneous aeronautical systems are rather closed and have been developed independently from terrestrial networks~\cite{azari2021evolution}.
Emerging technology concepts such as in-network computing and Software-Defined Networking (SDN) can provide a basis for a full integration of aeronautical systems to terrestrial communication.
The big challenge lies in the necessary management and orchestration approaches to address and control the resources of all participating aeronautical components in the sky forming a unified three-dimensional 6G system. 

In this article, we sketch a 6G aeronautical network architecture with a novel middle layer between satellite networks and terrestrial networks.
We regard the communication, processing and storage resources of all systems in the sky as the key elements of our architecture.
To leverage these resources, we advocate for a federation framework for the collection, monitoring and distribution of these resources among the heterogeneous flying objects to offer end-to-end high-performance services for new aeronautical applications.
Our description of the envisaged federation framework is complemented by the elaboration of three use cases including \textbf{Aeronautical Edge Computing (AEC)}, \textbf{aircraft-as-a-sensor}, and \textbf{Aircraft Cabin Networks}.   
We further use those scenarios to derive  research  challenges  for  6G  aeronautical  networks.

%
%
\section{6G Aeronautical Network Architecture}

\begin{figure}[t]
	\centering
	\captionsetup{font=small}
	\includegraphics[width=1\columnwidth]{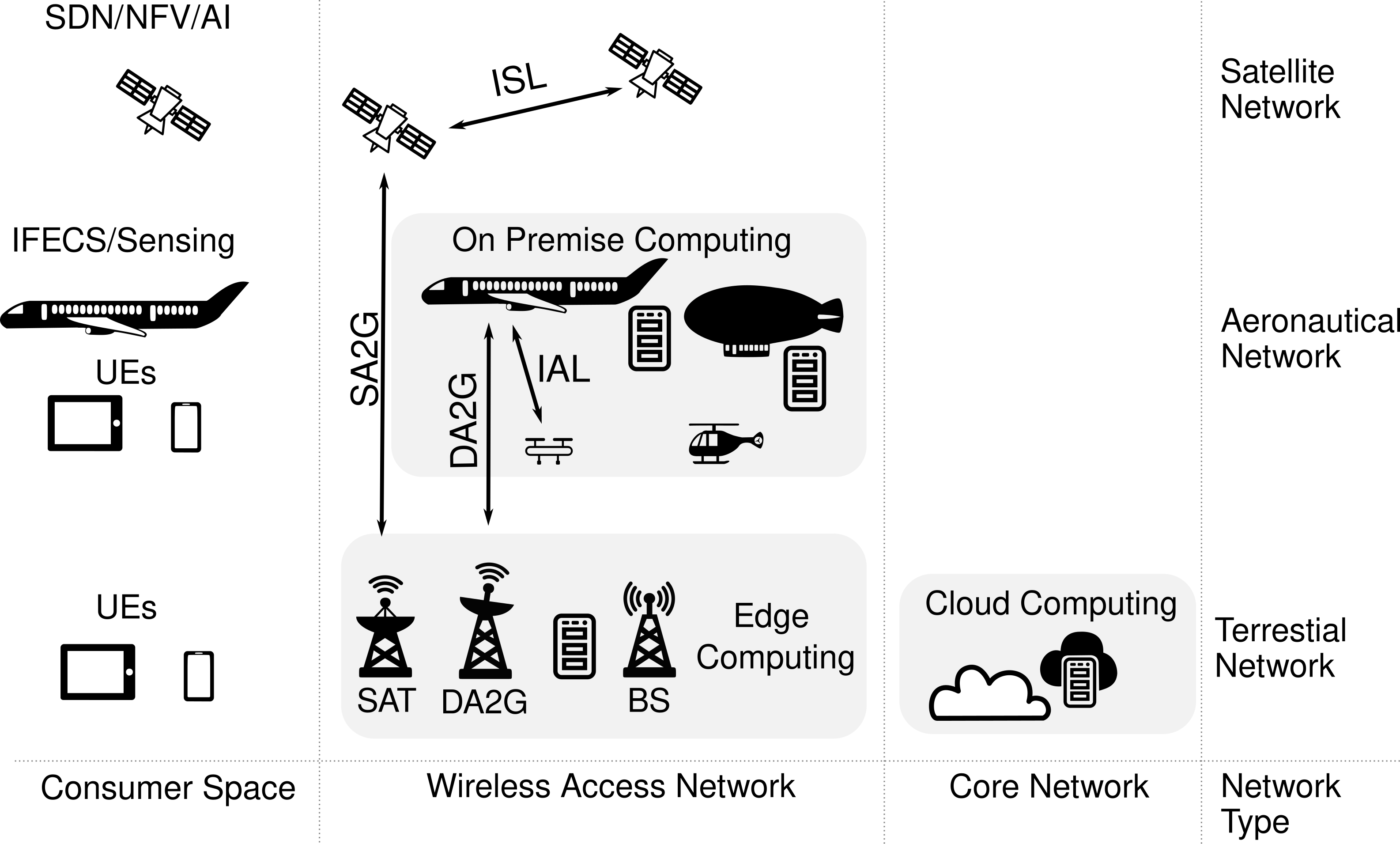}
	\caption{Envisioned 6G network architecture, including terrestrial, satellite and aeronautical networks.
		The satellite and aeronautical networks connect to the core network through the wireless access network.
		The consumer space consists of satellites, aircrafts and UEs. Resources consist of communication, computing and storage.}
	\label{fig:architecture}
\end{figure}

Next generation 6G networks anticipate the connection of everything, leading to the merging of terrestrial with non-terrestrial networks.
However, as pointed out in~\cite{huang2019airplane}, a middle layer between the satellite and terrestrial layers, namely the aeronautical layer is required in order to provide improved resource sharing and spectral efficiency.
Consequently, civil aircrafts, UAVs and HAPs will be utilized in combination with mm-wave links to provide high bandwidth.
This envisions a three-dimensional network architecture, which includes terrestrial i.e., Radio Access Networks (RANs), non-terrestrial i.e., satellites and aeronautical networks as elaborated in~\cite{dao2021survey, azari2021evolution}.
A detailed overview of the envisioned architecture is portrayed in Fig.~\ref{fig:architecture}.

Different from previous network generations and current research for aeronautical systems, we introduce a new concept for the {\it network consumer}.
In our perspective, a consumer could be traditional User Equipment (UEs), as well as UEs within the aircrafts requesting In-Flight Entertainment and Connectivity Services (IFECS).
Furthermore, aircrafts themselves requiring for instance weather predictions, sensing and localization data could be considered consumers.
In turn, traditional satellites may envision processing capacity or intelligence to run sophisticated algorithms which require SDN, Network Function Virtualization (NFV) and AI capabilities.
For the remainder of this paper, we will refer to the services provided to consumers as \textbf{applications}.

In order to map 6G features to our architecture, key elements are \textbf{communication}, \textbf{processing} and \textbf{storage} resources of all components in the sky.
Not only terrestrial networks, but also the aeronautical networks shall be equipped with servers that can in turn be utilized to improve the network performance, which advocate the concept of {\it federated in-network computing}.
Regarding the vicinity to the consumer, they are classified from cloud to edge and premise computing.
These servers constitute the \textbf{processing} and \textbf{storage} resources.
The \textbf{communication} resources in the sky consist of Inter-Aerial Links (IALs) for the communication among various aeronautical components, Inter-Satellite Links (ISLs) for the communication among satellites, Direct-Air-to-Ground (DA2G) links and Satellite-to-Gateway links (SA2G).
To establish the integration of the satellite and aeronautical components to the terrestrial network, Satellite Gateways (SATs), DA2G gateways and Base Stations (BSs) are placed in what we call \textbf{wireless access network}.
The wireless access network further connects to the \textbf{core network}.
In the latter, we will elaborate how resources can be gathered and allocated to consumers to provide the required applications through a federation entity in the sky.

\section{Related work}

In the last couple of years, satellite and aeronautical networks have attracted a lot of attention to establish network ubiquity.
In the context of 6G communication, aeronautical networks play a major role in complementing terrestrial networks~\cite{azari2021evolution}.
It is even suggested that several components of the aeronautical networks shall be enhanced with mobile radio capabilities.
For instance, enhancing HAPs and UAVs operating as flying BSs~\cite{strinati20206g} in order to serve UEs ubiquitously and in a timely manner.
With so much processing demand it is of utmost importance to build a highly reliable and effective management and orchestration entity in the sky to complement a similar one on the ground.

Moreover, the aeronautical network can be utilized as a surveillance system, which can alleviate the resource allocation and Virtual Network Function (VNF) placement for different services as proposed in~\cite{sun2020surveillance}.
Likewise, the concepts of AI can enhance network capabilities as presented in~\cite{yang2020artificial}.

In a similar fashion, satellite networks are gaining high traction in recent years, which has led to the development of methods to bridge the air and ground networks.
As a result, the concept of SDN has also been incorporated~\cite{papa2020design}.
Given all the aforementioned concepts anticipated for satellites, the requirement for high computing entities becomes mandatory.
While providing those services to satellites from the ground network is possible, it introduces extremely high and non-tolerable delays for 5G networks (i.e., $>$ $10$\,ms).
To avoid such issues, MEC-enabled sky solutions are suggested.
In this work, we stress the importance of such a component in the sky and shed light on a framework that fully establishes these requirements, while detailing the main elements and interfaces that allow for a normal operation and integration with other network system components.

A similar approach for a collaborative MEC system based on air and ground is presented in~\cite{qin2021air} and~\cite{cheng2018air}, which are the closest to our proposal.
However, as the majority of the state-of-the-art, they both consider MEC establishment mainly for the service of UEs and enhancing their performance.
Furthermore, they do not provide clear instructions with respect to the MEC operation, interaction among components as well as interfaces for UEs or third parties to collaborate with.
In contrast to already existing work, in this paper we focus explicitly on proposing clear instructions for the interaction among functions as well as interfaces compliant to 3GPP standardization.
In that regard, we demystify in depth the use-case of edge computing in the sky (i.e., AEC) and demonstrate how different vendors and actors in the sky can benefit from such a structure.

\section{Enabling 6G Applications in the Sky: Aeronautical Federation Framework}
\label{sec:federation}

To group available resources and create an infrastructure to collect and distribute them accordingly when needed, we propose a \textbf{federation framework} for management and orchestration.
The location of the federation framework could be any of the aeronautical components with adequate processing capabilities, for instance the federation framework could be established in a powerful HAP or specifically designed aircrafts.
The main goal is to group computing, storage and communication resources from infrastructure providers (i.e., satellite network providers, aircraft manufacturers, storage and computing providers).
Furthermore, with the collected resources, an intelligent SDN controller is responsible for the optimal allocation of the aforementioned resources to the incoming applications.
In the latter, we provide a detailed description of the management and orchestration procedures.

\subsection{Management and Orchestration}

The main components of the \textbf{federation framework} are depicted in Fig.~\ref{fig:federation_block}.
Each component provides a unique function through the interaction with other components.
An elaborated explanation is provided as follows:

\begin{figure}[t]
	\centering
	\captionsetup{font=small}
	\includegraphics[width=1\columnwidth, trim={0cm 0cm 0cm 0cm},clip]{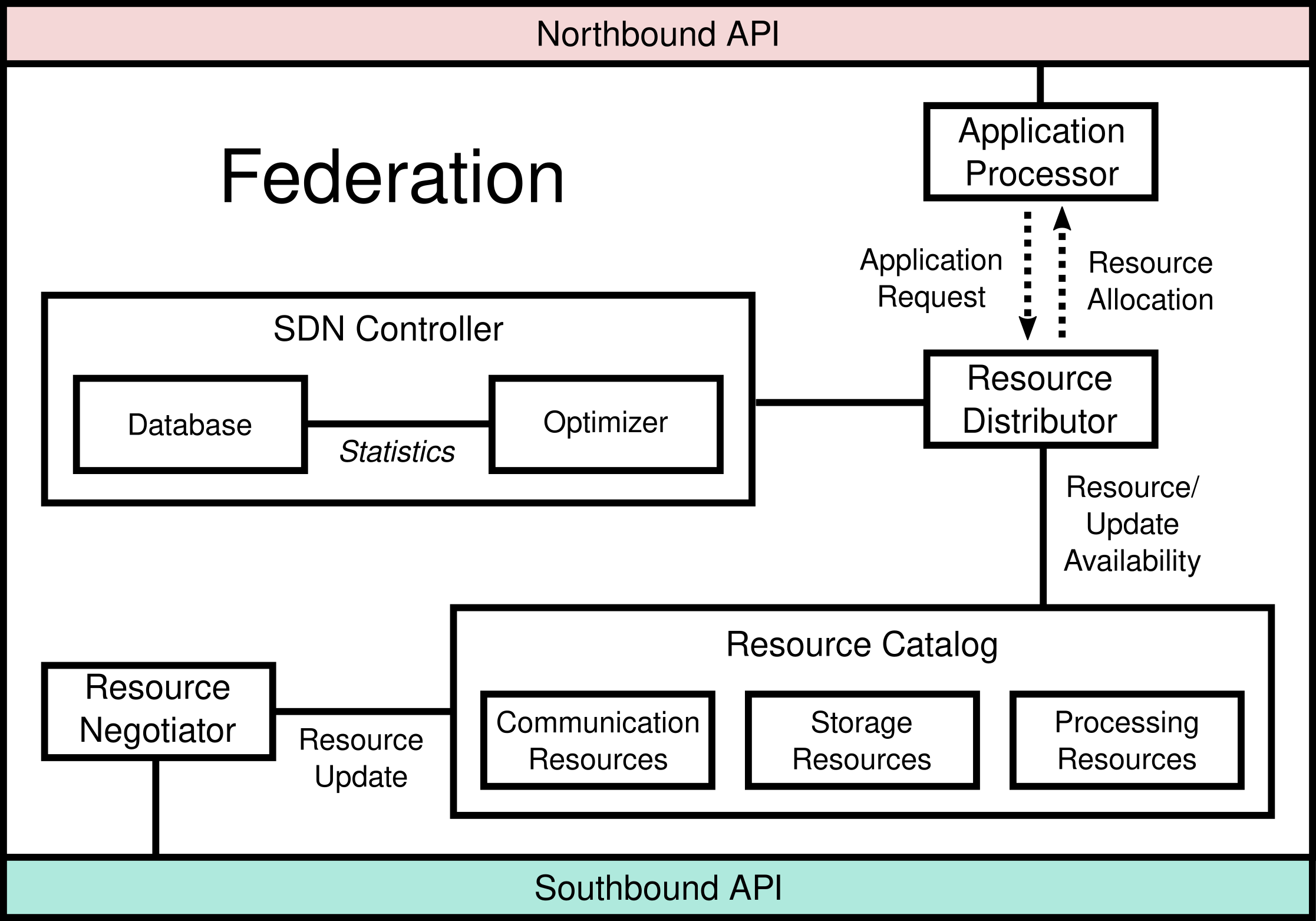}
	\caption{Overview of the federation framework. It provides details of the blocks that compose the federation, as well as the interaction among them.}
	\label{fig:federation_block}
\end{figure}

\begin{itemize}
	\item[1.] \textbf{Application Processor:} function responsible for the communication with the application layer.
	Through the northbound Application Programming Interface (API), the application processor obtains all application requests and forwards them to the federation framework.
	\item[2.] \textbf{SDN Controller:} is the heart of the federation framework.
	The SDN controller manages and distributes all system resources to the respective applications, regarding their needs.
	It is composed of a database to store information with respect to applications (i.e., resource utilization, network characteristics) as well as an optimization block that performs optimal resource allocation.
	\item[3.] \textbf{Resource Catalog:} as the name suggests it contains and preserves all the resources of the federation framework.
	\item[3a.] \textbf{Processing Resources:} resources that concern processing capabilities.
	This can range from cloud computing to edge and premise computing.
	\item[ 3b.] \textbf{Storage Resources:} resources that concern content storage.
	For instance caching and servers.
	\item[ 3c.] \textbf{Communication Resources:} resources that deal with the communication aspect.
	These resources can consist of satellite links, aeronautical links, satellite-ground links or radio access terrestrial network links.
	\setlength{\itemindent}{0em}\item[4.] \textbf{Resource Negotiator:} responsible for negotiating and obtaining resources from the infrastructure providers.
	The goal of this block is to secure resources in a smart manner in order to maximize profit.
	This process is realized through the southbound API.
	\setlength{\itemindent}{0em}\item[5.] \textbf{Resource Distributor:} function responsible for the communication with the resource catalog to identify available resources.
	It further forwards resource information to the SDN controller and finally distributes resources to applications.
\end{itemize}

\begin{figure}[t]
	\centering
	\captionsetup{font=small}
	\includegraphics[width=1\columnwidth, trim={0cm 21cm 8cm 0cm},clip]{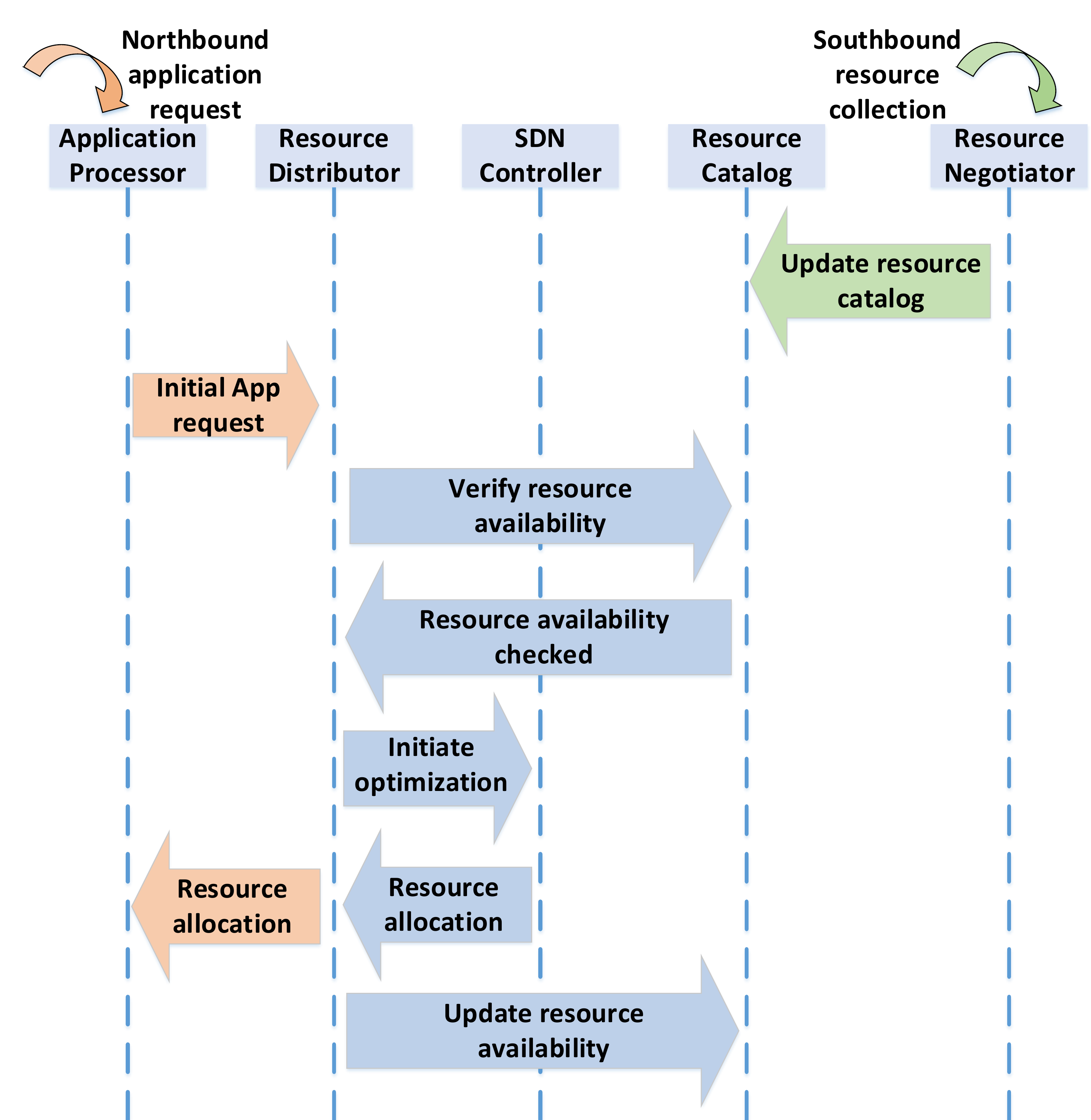}
	\caption{Flow diagram of the interaction among the federation framework components during an application request.}
	\label{fig:federation_block_diagram}
\end{figure}

\subsection{Federation Framework Block Interaction}
\label{subsec:federation}

To better highlight the communication among all federation blocks, and to detail the exchanged information, a flow chart is presented in Fig.~\ref{fig:federation_block_diagram}.

According to Fig.~\ref{fig:federation_block_diagram}, the federation framework communicates with other entities i.e., applications, services and infrastructure providers while using APIs.
We refer to these APIs as northbound for the communication with the applications and southbound for the collection of resources, accordingly.
The APIs include the 5G information model known as Network Resource Model (NRM) based on Extensible Markup Language (XML) or JavaScript Object Notation (JSON) formats.

Initially, the \textbf{resource negotiator} acquires the computing, storage and communication resources from the service and infrastructure providers through the southbound API.
This serves as an intermediary, where several algorithms can be utilized to obtain this goal, e.g., game theory.
Once the resources have been secured, a connection with the \textbf{resource catalog} is established to update the resource availability.
Equivalently, applications request resources from the federation framework using the northbound API through the \textbf{application processor}.
Once the request has been gathered and processed by the \textbf{application processor}, the flow of the data is transferred to the \textbf{resource distributor}.
The \textbf{resource distributor} in turn communicates with the \textbf{resource catalog} to verify resource availability.
If resources are available, a positive feedback is reported to the \textbf{resource distributor}.
Now, the \textbf{SDN controller} block is triggered, which is in charge of optimal resource allocation.
In this block, several optimization algorithms can be deployed, including recent advanced AI solutions~\cite{yang2020artificial}.
Once the optimal decision is acquired, the \textbf{SDN controller} forwards the information of resource allocation to the \textbf{resource distributor}.
Finally, the decision is made available to the applications through the \textbf{application processor}.

%
%
%
\section{6G Aeronautical Applications}
The proposed federation framework enables new aeronautical use cases that benefit from a full integration of aeronautical communication and services in the 6G ecosystem.
In this section, we discuss three use cases in particular, while demystifying how the federation framework alleviates their functionality.

\subsection{Aeronautical Edge Computing (AEC)}

As aforementioned, satellite networks are gaining increased attention in the 5G/6G era, where concepts such as SDN, AI and NFV are envisioned to be incorporated into them.
However, given the limited hardware capabilities that satellites possess i.e., low processing units and memory, such applications become difficult to be executed.
Furthermore, aeronautical elements such as UAVs and HAPs are anticipated to serve as flying BSs and provide computing capabilities to UEs.
In turn, already existing IFECS are urging for maintaining emerging delay and throughput critical applications to passengers.
To this end, existing non-terrestrial and aeronautical network infrastructures relying solely on the terrestrial networks and core network servers to achieve their services become obsolete.

To overcome the aforementioned issues, we envision an AEC solution, which we make available through the proposed aeronautical federation framework.
In our proposal, aeronautical components serve as flying computing entities.
This solution, not only maintains low latency for emerging 6G applications in the sky, but also increases the importance of non-terrestrial and aeronautical systems in the 6G era.

\begin{figure}[t]
	\centering
	\captionsetup{font=small}
	\includegraphics[width=1\columnwidth, trim={0cm 0cm 0cm 0cm},clip]{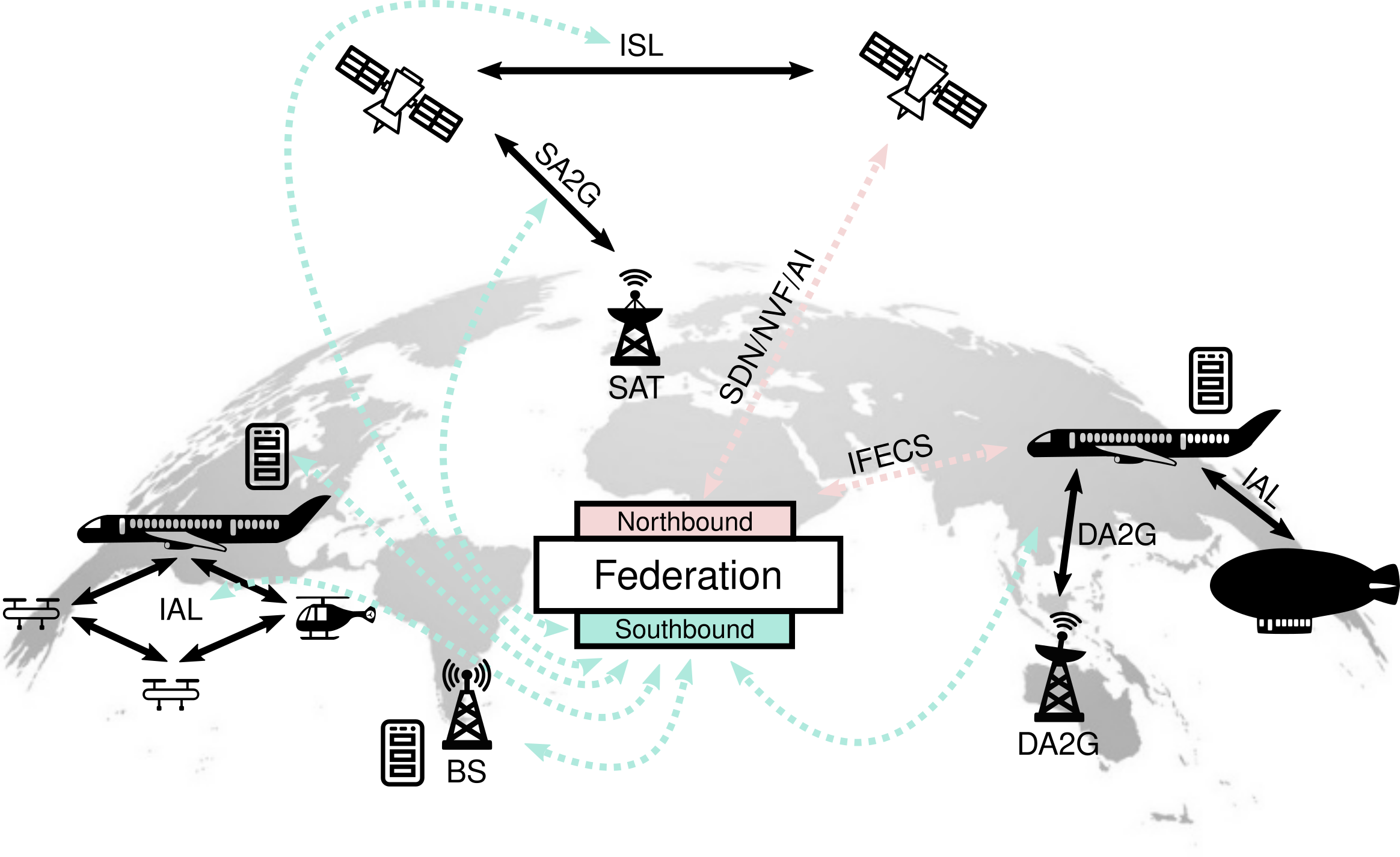}
	\caption{Envisioned 6G AEC use case through the proposed aeronautical federation framework.
		Processing, storage and communication resources marked with green are acquired by the infrastructure provides through the southbound API.
		Application requests, marked with red are forwarded to the federation framework through the northbound API.}
	\label{fig:mac}
\end{figure}

An overview of the proposed approach is demonstrated in Fig.~\ref{fig:mac}.
Here we portray how the aeronautical network can be enhanced with computing entities, which can be utilized to always follow flight paths and satellites serving them.
In the figure, IALs, ISLs, DA2G links and SA2G links form the \textbf{communication resources}.
Drones, HAPs and aircrafts with storage and processing capabilities constitute the \textbf{storage} and \textbf{processing} resources respectively.
These resources, portrayed with green color in Fig.~\ref{fig:mac}, are acquired by the federation framework using the southbound API through the resource negotiator block as explained in Section~\ref{subsec:federation}.

SDN/NFV and AI-enabled satellites, UEs within the aircraft (i.e., IFECS) as well as UEs on the ground compose the \textbf{applications}.
Application requests, marked with red in Fig.~\ref{fig:mac}, are forwarded through the northbound API to the federation framework and are processed from the \textbf{application processor}.
Once the optimal solution is provided by the \textbf{SDN controller}, the application requirements are established.
Therefore, the necessity to acquire resources from the terrestrial network is eliminated and low latency is achieved. 

\begin{figure}[t]
	\centering
	\captionsetup{font=small}
	\includegraphics[width=0.48\textwidth]{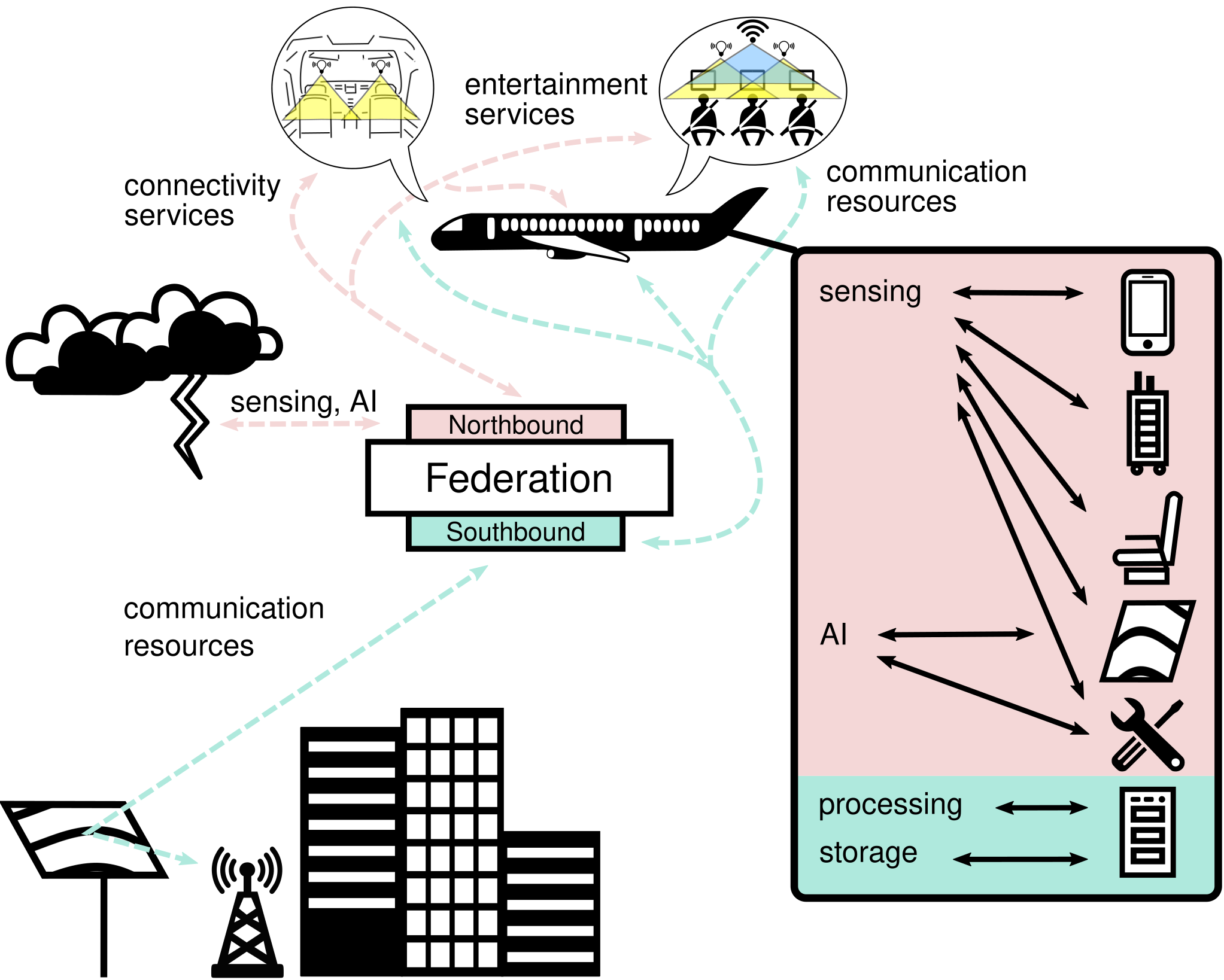}
	\caption{Overview of in and outside aircraft applications enabled through the aeronautical federation framework.
		Storage and processing resources are marked with green, whereas sensing, AI-enabled, entertainment and connectivity applications are labeled with red.}
	\label{fig:applications}
\end{figure}

\subsection{Aeronautical Sensor Networks}

6G holds the potential for different applications of sensing inside and outside an aircraft.
Inside an aircraft, for instance, it may be used to localize passengers with an expected centimeter resolution via their mobile devices and seat screens.
This allows an automated and accurate identification of passengers enabling new passenger seat centered services, such as targeted advertisements or differential customer experience depending on the passenger category.
Moreover, UE localization in the aircraft fosters an automated installation of 6G-enabled seat hardware, compared to the current manual configuration.

The envisaged 6G AI capability will allow to predictively deliver data to individual in-cabin elements, such as seat screens of the entertainment system or passenger devices.
The sensing feature could also be used in conjunction with the AI capability to perform predictive maintenance and anticipate which parts of the connected aircraft are at the risk of failure and thus need proactive replacement. 
Additionally, the sensing capability can be combined with the usage of intelligent surfaces~\cite{9183763}.
This can improve spectral efficiency inside the aircraft by increasing the spectral diversity in Multi-Input Multi-Output (MIMO) communication or by focusing the beam in beamforming resulting in a higher throughput. 

Finally, sensing combined with 6G's context awareness informs applications about the ideal time to communicate \cite{bourdoux20206g}.
Thereby reducing the energy consumption of transmissions and contributing to the energy efficiency of the aircraft's network.
All aforementioned use cases benefit from the federation framework proposed in this work.
Furthermore, the federation framework also enables applications outside an aircraft using 6G’s sensing and AI capabilities. 
Specifically, the collected weather data by an aircraft can be used to perform better channel estimations. 
Channel estimations, in turn, could also be used to improve the weather forecast. 
An illustration for a sensing and weather prediction scenario utilizing the aeronautical federation framework is depicted in Fig.~\ref{fig:applications}.
In this setup, processing and storage resources marked as green are acquired by the federation framework through the southbound interface.
Applications labeled with red, consist of in-cabin sensing features and AI-based weather forecast.
The applications' requests are forwarded to the federation framework through the northbound API.
The built-in SDN controller and optimizer is used to optimally allocate resources for the multiple access points found in an aircraft. 
It also coordinates the configuration of the intelligent surfaces to broadcast efficiently weather prediction data to avoid Non-Line of Sight (NLoS) and bad wireless channel conditions.

\subsection{Aircraft Cabin Networks}

In-cabin communication applications, such as entertainment services, video streaming, passenger communication, crew and cockpit communication, and passenger health and safety monitoring, have diverse QoS requirements. 
For example, crew and cockpit communications have a stricter latency and reliability requirement compared to all aforementioned applications, especially due to the security and safety concerns.
In turn, IFECS demand a high bandwidth. Multiple wireless technologies cooperating together are envisioned to satisfy these extreme performance requirements in 6G networks.

The key challenge in enabling such cooperation is to optimize the resource allocation of multiple access technologies considering UE requirements, environment mobility, network constraints, and failovers. Such optimization, in a multi-technology environment, requires a framework connecting different network components for seamless interaction. 
The federation framework introduced in this work functions as an enabler for resource management and orchestration of an in-cabin network with various wireless access technologies. 

This federation framework pools together all communication resources and hence can allocate them across multiple access technologies as depicted in Fig.~\ref{fig:applications}. 
This enables the aggregation of wireless links of the various wireless technologies which is executed by multi-path transport protocols like Multi Path Transport Control Protocol (MPTCP)~\cite{rfc6824}. 
Link aggregation also has the advantage of enabling seamless and flexible switching between the different technologies in case of a link failure. Link aggregation can be best utilized by access technologies that do not interfere with each other. 
In recent years, Light-Fidelity (LiFi) has emerged as a promising technology that can be integrated into aircraft cabin networks. 
This envisioned in-aircraft LiFi-RF network is depicted in Fig.~\ref{fig:applications}. Since MPTCP combines the technologies on the transport layer without changing the underlying network infrastructure, it requires no major changes to the network architecture or user devices thus enabling easy integration with existing infrastructure.

%
\section{Initial System Validation}

In order to discuss the validity of the presented concept of the aeronautical federation framework described in Section~\ref{sec:federation}, we focus on the Aeronautical Edge Computing (AEC) use case. 
In our previous work~\cite{papa2021cost} we describe a prototype of this concept. 
We envision the aeronautical federation framework to be established within an aircraft. 
As in our proposed concept, the provisioning of services is extended across the sky, where UEs in other aircrafts could be additionally served to extend the operational region of the proposed framework.

To illustrate the benefits of the federation framework, we show some results based on the optimization problem described in~\cite{papa2021cost}. 
Here, we maximize the number of served \textbf{network slices} in an aircraft by optimally allocating caching and satellite link resources. 
We refer to a network slice as a group of UEs that share similar characteristics (e.g., request the same video files) and that have the same delay requirements.
For the evaluation of our federation engine concept, we consider two approaches. 
The first approach utilizes both caching resources (i.e., storage resources) and satellites links (i.e., communication resources) to serve UEs within aircrafts demonstrating the \textbf{AEC} feature. 
In contrast, the second approach only utilizes satellite resources (i.e., current network infrastructures). 
The latter serves as our \textbf{baseline}.

\begin{figure}[t]
	\centering
	\captionsetup{font=small}
	\includegraphics[width=0.4\textwidth]{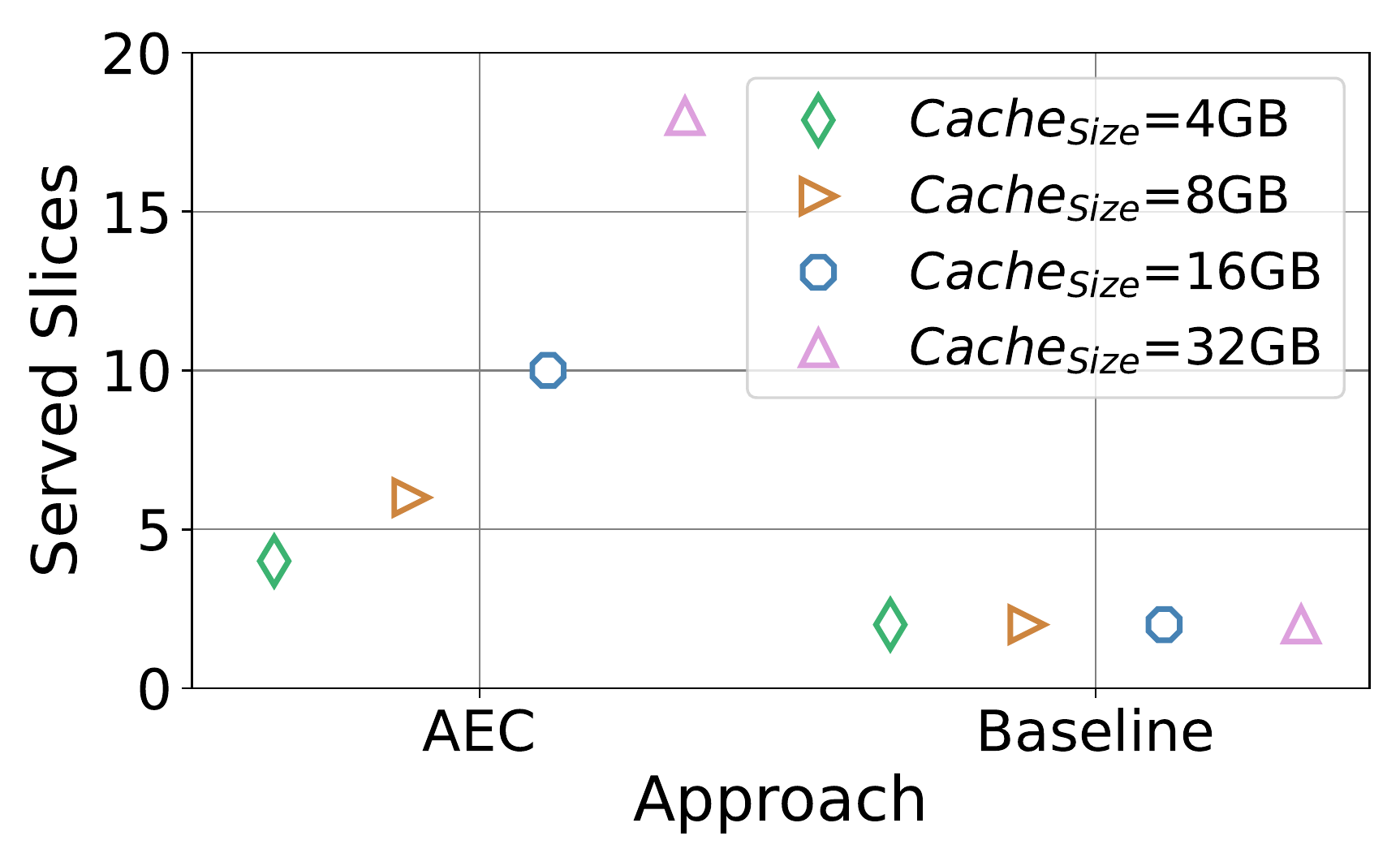}
	\caption{AEC scenario compared to baseline approach with no caching for various cache sizes in terms of number of served slices inside an aircraft.}
	\label{fig:validation}
\end{figure}

For an initial validation, we consider the following parameters: 
a Geostationary Earth Orbit (GEO) satellite with $250$\,ms round-trip delay to the core network in the ground and $112$\,Mbps bandwidth that is shared among UEs in the aircraft. 
In turn, UEs are grouped into network slices, where each network slice contains 15 UEs. 
Each UE within a network slice requests video files from a catalog $\mathcal{K}$ of 10000 files. 
The arrival rate per UE is $2$ files/s and each file is of size $0.2$\,MB. 
In total there are $30$ network slices, resulting in a necessary storage of $60$\,GB. 
For the sake of simplicity and due to space limitations, we assume that each network slice has the same delay requirement of $1$\,s and its UEs request video files following a zipf 
distribution with an exponent of $0$ (i.e., uniform distribution of files in $\mathcal{K}$). To represent the AEC scenario, inside the aircraft a cache is shared among network slices. 
Results for various cache sizes are portrayed in Fig.~\ref{fig:validation} and are compared to the baseline approach (i.e., no caching) with respect to the number of served slices.

According to the results illustrated in Fig.~\ref{fig:validation}, the AEC approach outperforms the baseline approach, $4$ compared to $2$ served slices for a cache size of $4$\,GB (i.e., $\approx 7$\% of the total needed storage). 
However the difference between the approaches increases with increasing cache size with $18$ served slices compared to $2$ for a cache size of $32$\,GB (i.e., $\approx 50$\% of the total needed storage). 
The rationale behind these results lies on the high round-trip delay introduced by satellites that hinder the service of additional slices. 
Moreover, since the baseline approach does not consider caching resources, increasing the cache size does not benefit the performance.
Thus, we can conclude that the utilizing caching inside an aircraft is more beneficial compared to the case when video files are retrieved from the core network, as the round-trip delay of the satellites is avoided.

\section{Discussion and Conclusion}
In this work we discuss the potentials for aeronautical applications in the 6G era.
In that regard, we emphasize the importance of a full integration of aeronautical systems and networks into the 6G terrestrial network to form a three-dimensional 6G network in the sky.
Therefore, we propose to leverage the resources of aeronautical systems to form an integrated federated system to provide end-to-end communication services.
Our proposed federation framework allows to utilize communication, processing and storage resources across different entities.
This concept is envisioned to provide better network usage and reduce the network latency and increase reliability while enabling emerging technologies such as AI and SDN in the sky.
Furthermore, we highlight three specific 6G-enabled aeronautical use cases and elaborate potential research directions.
For the use case of AEC, we demonstrate with an example how our proposed framework can reduce network latency and thus serve more network slices compared to traditional network architectures in the sky that utilized only satellite resources.

\section*{Acknowledgement}
The authors acknowledge the financial support by the Federal Ministry of Education and Research of Germany (BMBF) in the programme of "Souverän. Digital. Vernetzt." joint project 6G-life, project identification number 16KISK002.

%

\bibliography{bibliography}{}

\begin{thebibliography}{10}
\providecommand{\url}[1]{#1}
\csname url@samestyle\endcsname
\providecommand{\newblock}{\relax}
\providecommand{\bibinfo}[2]{#2}
\providecommand{\BIBentrySTDinterwordspacing}{\spaceskip=0pt\relax}
\providecommand{\BIBentryALTinterwordstretchfactor}{4}
\providecommand{\BIBentryALTinterwordspacing}{\spaceskip=\fontdimen2\font plus
\BIBentryALTinterwordstretchfactor\fontdimen3\font minus
  \fontdimen4\font\relax}
\providecommand{\BIBforeignlanguage}[2]{{%
\expandafter\ifx\csname l@#1\endcsname\relax
\typeout{** WARNING: IEEEtran.bst: No hyphenation pattern has been}%
\typeout{** loaded for the language `#1'. Using the pattern for}%
\typeout{** the default language instead.}%
\else
\language=\csname l@#1\endcsname
\fi
#2}}
\providecommand{\BIBdecl}{\relax}
\BIBdecl

\bibitem{azari2021evolution}
M.~M. Azari, S.~Solanki, S.~Chatzinotas, O.~Kodheli, H.~Sallouha, A.~Colpaert,
  J.~F.~M. Montoya, S.~Pollin, A.~Haqiqatnejad, A.~Mostaani \emph{et~al.},
  ``{Evolution of Non-Terrestrial Networks From 5G to 6G: A Survey},''
  \emph{arXiv preprint arXiv:2107.06881}, 2021.

\bibitem{strinati20206g}
E.~C. Strinati, S.~Barbarossa, T.~Choi, A.~Pietrabissa, A.~Giuseppi,
  E.~De~Santis, J.~Vidal, Z.~Becvar, T.~Haustein, N.~Cassiau \emph{et~al.},
  ``{6G in the sky: On-demand Intelligence at the edge of 3D networks},''
  \emph{arXiv preprint arXiv:2010.09463}, 2020.

\bibitem{henry20205g}
S.~Henry, A.~Alsohaily, and E.~S. Sousa, ``{5G is Real: Evaluating the
  Compliance of the 3GPP 5G New Radio System with the ITU IMT-2020
  Requirements},'' \emph{IEEE Access}, vol.~8, pp. 42\,828--42\,840, 2020.

\bibitem{viswanathan2020communications}
H.~Viswanathan and P.~E. Mogensen, ``{Communications in the 6G era},''
  \emph{IEEE Access}, vol.~8, pp. 57\,063--57\,074, 2020.

\bibitem{huang2019airplane}
X.~Huang, J.~A. Zhang, R.~P. Liu, Y.~J. Guo, and L.~Hanzo, ``{Airplane-Aided
  Integrated Networking for 6G Wireless: Will it work?}'' \emph{IEEE Vehicular
  Technology Magazine}, vol.~14, no.~3, pp. 84--91, 2019.

\bibitem{dao2021survey}
N.-N. Dao, Q.-V. Pham, N.~H. Tu, T.~T. Thanh, V.~N.~Q. Bao, D.~S. Lakew, and
  S.~Cho, ``{Survey on Aerial Radio Access Networks: Toward a Comprehensive 6G
  Access Infrastructure},'' \emph{IEEE Communications Surveys \& Tutorials},
  vol.~23, no.~2, pp. 1193--1225, 2021.

\bibitem{sun2020surveillance}
J.~Sun, F.~Liu, Y.~Zhou, G.~Gui, T.~Ohtsuki, S.~Guo, and F.~Adachi,
  ``{Surveillance Plane Aided Air-Ground Integrated Vehicular Networks:
  Architectures, Applications, and Potential},'' \emph{IEEE Wireless
  Communications}, vol.~27, no.~6, pp. 122--128, 2020.

\bibitem{yang2020artificial}
H.~Yang, A.~Alphones, Z.~Xiong, D.~Niyato, J.~Zhao, and K.~Wu,
  ``{Artificial-Intelligence-Enabled Intelligent 6G Networks},'' \emph{IEEE
  Network}, vol.~34, no.~6, pp. 272--280, 2020.

\bibitem{papa2020design}
A.~Papa, T.~De~Cola, P.~Vizarreta, M.~He, C.~Mas-Machuca, and W.~Kellerer,
  ``{Design and Evaluation of Reconfigurable SDN LEO constellations},''
  \emph{IEEE Transactions on Network and Service Management}, vol.~17, no.~3,
  pp. 1432--1445, 2020.

\bibitem{qin2021air}
Z.~Qin, H.~Wang, Y.~Qu, H.~Dai, and Z.~Wei, ``{Air-Ground Collaborative Mobile
  Edge Computing: Architecture, Challenges, and Opportunities},'' \emph{arXiv
  preprint arXiv:2101.07930}, 2021.

\bibitem{cheng2018air}
N.~Cheng, W.~Xu, W.~Shi, Y.~Zhou, N.~Lu, H.~Zhou, and X.~Shen, ``{Air-Ground
  Integrated Mobile Edge Networks: Architecture, Challenges, and
  Opportunities},'' \emph{IEEE Communications Magazine}, vol.~56, no.~8, pp.
  26--32, 2018.

\bibitem{9183763}
H.~Hashida, Y.~Kawamoto, and N.~Kato, ``{Intelligent Reflecting Surface
  Placement Optimization in Air-Ground Communication Networks Toward 6G},''
  \emph{IEEE Wireless Communications}, vol.~27, no.~6, pp. 146--151, 2020.

\bibitem{bourdoux20206g}
A.~Bourdoux, A.~N. Barreto, B.~van Liempd, C.~de~Lima, D.~Dardari, D.~Belot,
  E.-S. Lohan, G.~Seco-Granados, H.~Sarieddeen, H.~Wymeersch, J.~Suutala,
  J.~Saloranta, M.~Guillaud, M.~Isomursu, M.~Valkama, M.~R.~K. Aziz,
  R.~Berkvens, T.~Sanguanpuak, T.~Svensson, and Y.~Miao, ``{6G White Paper on
  Localization and Sensing},'' 2020.

\bibitem{rfc6824}
\BIBentryALTinterwordspacing
A.~Ford, C.~Raiciu, M.~J. Handley, and O.~Bonaventure, ``{TCP Extensions for
  Multipath Operation with Multiple Addresses},'' RFC 6824, Jan. 2013.
  [Online]. Available: \url{https://www.rfc-editor.org/info/rfc6824}
\BIBentrySTDinterwordspacing

\bibitem{papa2021cost}
A.~Papa, H.~M. G{\"u}rsu, L.~Goratti, T.~Rasheed, and W.~Kellerer, ``{Cost of
  Network Slice Collaboration: Edge Network Slicing for In-Flight
  Connectivity},'' in \emph{IEEE International Conference on Communications},
  2021, pp. 1--6.

\end{thebibliography}
\balance
\bibliographystyle{IEEEtran}

\end{document}